\documentclass[twocolumn,showpacs,preprintnumbers,
               superscriptaddress]{revtex4}

\usepackage{graphicx}

\newcommand{\<}{\langle}
\renewcommand{\>}{\rangle}
\newcommand{\Tr}{\mbox{\textrm{Tr}}}
\renewcommand{\Re}{\mbox{\textrm{Re}}}
\renewcommand{\Im}{\mbox{\textrm{Im}}}

\begin{document}

\preprint{HUPD-0112, YITP-01-61, UTCCP-P-111}

\title{Responses of hadrons to chemical potential at finite temperature}

\author{S. Choe}
\affiliation{Department of Physics, Hiroshima University,
             Higashi-Hiroshima 739-8526, Japan}

\author{Ph. de Forcrand}
\affiliation{Institut f\"ur Theoretische Physik, ETH-H\"onggerberg,
             CH-8093 Z\"urich, Switzerland}
\affiliation{Theory Division, CERN, CH-1211 Geneva 23, Switzerland}

\author{M. Garc\'{\i}a P\'erez}
\affiliation{Theory Division, CERN, CH-1211 Geneva 23, Switzerland}

\author{S. Hioki}
\affiliation{Department of Physics, Tezukayama University,
             Nara 631-8501, Japan}

\author{Y. Liu}
\affiliation{Department of Physics, Hiroshima University,
             Higashi-Hiroshima 739-8526, Japan}

\author{H. Matsufuru}
\affiliation{Yukawa Institute for Theoretical Physics,
             Kyoto University,  Kyoto 606-8502, Japan}

\author{O. Miyamura}
\affiliation{Department of Physics, Hiroshima University,
             Higashi-Hiroshima 739-8526, Japan}

\author{A. Nakamura}
\affiliation{IMC, Hiroshima University,
             Higashi-Hiroshima 739-8521, Japan}

\author{I.-O. Stamatescu}
\affiliation{Institut f\"ur Theoretische Physik, Universit\"at
             Heidelberg,  D-69120 Heidelberg, Germany}
\affiliation{FEST, Schmeilweg 5, D-69118 Heidelberg, Germany}

\author{T. Takaishi}
\affiliation{Hiroshima University of Economics,
             Hiroshima 731-0192, Japan}

\author{T. Umeda}
\affiliation{Center for Computational Physics, University of Tsukuba,
             Tsukuba 305-8577, Japan}

\collaboration{QCD-TARO Collaboration}


\date{July 8, 2001, revised \today}

\begin{abstract}
We present a framework to compute the responses of hadron masses
to the chemical potential in lattice QCD simulations.
As a first trial, the screening mass of the pseudoscalar meson and its
first and second responses are evaluated.
We present results on a $16\times 8^2\times 4$
lattice with two flavors of staggered quarks below and above $T_c$.
The responses to both the isoscalar and isovector chemical potentials
are obtained.
They show different behavior in the low and the high
temperature phases, which may be explained as a consequence of
chiral symmetry breaking and restoration, respectively.
\end{abstract}

\maketitle

\section{Introduction}

It is well known that studying finite density QCD through lattice
simulations is a very hard problem.
The fermionic determinant at finite chemical potential is complex,
and gives an oscillating behavior in quantum averages which makes
simulations very inefficient.
Since the naive quenched approximation at finite chemical
potential leads to an essentially different world \cite{Stephanov},
the use of dynamical fermions would be essential to extract the relevant
physics.
In spite of these difficulties, the study of hadrons in a finite
baryonic environment is quite important
\cite{Hatsuda-Lee,Hayashigaki,Agaki},
in view of recent experimental developments
and of the theoretical interest in the phase structure of QCD.
Search for the quark gluon plasma phase in  high energy heavy ion
collision experiments requires theoretical understanding
of hadronic properties at finite temperature and density \cite{NA50}.
Moreover, some experimental results can be interpreted by assuming a shift
in the mass and the width of the $\rho$ meson, induced by the dense nuclear
medium even below the deconfinement transition \cite{rapet}.

In this paper we propose a new technique to investigate non-zero
chemical potential using lattice QCD simulations.
There are several approaches to circumvent the difficulty of
studying a finite chemical potential system, and they seem
successful to a limited extent \cite{Glasgow,Cano,imag}.
In particular, the study of baryon number susceptibility
at zero baryon density has reported an abrupt jump at the
transition temperature \cite{Gott}.
There is in fact much interesting physical information which
can be extracted from the behavior of a system at small chemical potential.
Our strategy is to expand the hadronic quantities, such as masses,
in the vicinity of zero chemical potential at finite temperature,
and explore their changes through the response to the chemical
potential at $\mu=0$.
This allows to perform the numerical simulations with standard
methods.
The Taylor expansion in $\mu$ has also been used, and its properties
discussed, in \cite{imag}. There, the
behavior of observables, measured by standard methods as a function of
an imaginary chemical potential $\mu$, is fitted by a Taylor series
amenable to analytic continuation to real $\mu$ (assuming a large enough
analyticity domain). Here we measure the Taylor coefficients directly, in a
single simulation, at $\mu=0$, by calculating the derivatives of the relevant
observables.
Although the Taylor expansion cannot reproduce the non-analyticity
inherent to a phase transition, it may suffice for observing the
rounded, analytic behavior indicative of a phase transition in a
finite volume, and for applying finite size scaling to probe
the transition in the infinite volume limit.
Our preliminary results have been reported in \cite{TARO1}.

This paper is organized as follows.
In the next section, we develop the basic formulae to evaluate the
first and the second responses of hadron screening masses with respect
to both of the isoscalar and the isovector chemical potentials.
In Section~\ref{sec3}, we report on Monte Carlo simulations performed
with two dynamical flavors of staggered quarks.
We obtain the responses of the pseudoscalar meson mass,
and discuss some implications of our results.
Section~\ref{sec4} presents our conclusions.

\section{Chemical potential response of hadron masses
         at finite temperature}
\label{sec2}

This section develops the basic framework to observe the response of
hadron masses with respect to the chemical potential.
At fixed temperature $T$ and bare quark masses,
the screening mass of a hadron is expanded in the form:
\begin{eqnarray}
 \left.\frac{M(\mu)}{T}\right|_{\mu}
 &=&  \left. \frac{M}{T}\right|_{\mu=0} + \left(\frac{\mu}{T}\right)
      \left(\frac{d M}{d\mu}\right)_{\mu=0}  \nonumber \\
 & & \hspace{-1.2cm}
      + \left(\frac{\mu}{T}\right)^2 \frac{T}{2}
       \left(\frac{d^2 M}{d\mu^2}\right)_{\mu =0}
 + O \left[ \left(\frac{\mu}{T}\right)^3 \right].
 \label{exp}
\end{eqnarray}
Assume that the spatial hadron correlator $C(x)$ is dominated by
a single pole contribution,
\begin{eqnarray}
 C(x)
  &\equiv&  \sum_{y,z,t}\<H(x,y,z,t)H(0,0,0,0)^{\dagger}\>
   \nonumber \\
  &=& A ( e^{-\hat{M}\hat{x}} +e^{-\hat{M}(L_x-\hat{x})} ) ,
\label{singlepole}
\end{eqnarray}
where $\hat{M}=aM$ and $\hat{x}=x/a$.
$L_x$ is the lattice size in the $x$-direction.
Generalization to a multi-pole situation is straightforward.
In the case of a local hadron operator,
$A$ is represented as $A=\hat{\gamma}/2\hat{M}$ with
the residue of the propagator, $\hat{\gamma}$.
We employ a smeared hadron operator $H(x,y,z,t)$ to enhance
the overlap with the ground state, and hence $A$ depends on the
choice of sources.
With a fixed type of smearing function, however,
the behavior of $A$ as a function of the chemical potential
provides information on the coupling to the medium.

We take the first and second derivatives of the hadron correlator
with respect to $\hat{\mu}\equiv a \mu = \mu/(N_t T)$,
where $\mu$ is the chemical potential.
While we here write down expressions which depend on a single chemical
potential, it is straightforward to generalize to the case
involving two flavors corresponding to $u$ and $d$ quarks.
Later we introduce the isovector and the isoscalar chemical
potential by setting $\mu_u$ and $\mu_d$ appropriately.
The first and the second derivatives of $C(x)$ with respect to
the chemical potential lead to
\begin{eqnarray}
\frac{1}{C(x)} \frac{dC(x)}{d\hat{\mu}}
 &=& \frac{1}{A} \frac{d A}{d\hat{\mu}}
 \nonumber \\
 & & \hspace{-2.6cm}
 + \frac{d\hat{M}}{d\hat{\mu}}
  \left\{ \left( \hat{x}-\frac{L_x}{2} \right) \tanh
     \left[ \hat{M} \left( \hat{x}-\frac{L_x}{2} \right) \right]
    - \frac{L_x}{2} \right\} ,
\label{deri-1}
\end{eqnarray}
and
\begin{widetext}
\begin{eqnarray}
\frac{1}{C(x)} \frac{d^2 C(x)}{d \hat{\mu}^2}
 &=&  \frac{1}{A} \frac{d^2 A}{d\hat{\mu}^2}
     + \left( \frac{2}{A} \frac{d A}{d\hat{\mu}}
          \frac{d \hat{M}}{d\hat{\mu}} +
         \frac{d^2 \hat{M}}{d\hat{\mu}^2} \right)
      \left\{ \left( \hat{x}-\frac{L_x}{2} \right) \tanh
             \left[ \hat{M} \left( \hat{x} - \frac{L_x}{2} \right)
         \right]  - \frac{L_x}{2} \right\}
 \nonumber \\
 & &
  + \left( \frac{d \hat{M}}{d\hat{\mu}} \right)^2
   \left\{  \left( \hat{x}-\frac{L_x}{2} \right)^2
       + \frac{L_x^2}{4} - L_x \left( \hat{x} - \frac{L_x}{2} \right)
         \tanh \left[ \hat{M} \left( \hat{x} - \frac{L_x}{2} \right)
    \right] \right\} .
\label{deri-2}
\end{eqnarray}
\end{widetext}
$C(x)$ and the derivatives of $C(x)$ are calculated from
lattice simulations.
Then, using the right-hand side of Eqs.~(\ref{deri-1}) and
(\ref{deri-2}), the first and the
second responses of hadron masses and couplings are determined.

The next step is to relate the derivatives of the correlator
to quantities which are measured in the lattice simulations.
In this work, we consider the flavor non-singlet mesons in QCD with two
flavors.
The hadron correlator is then given by
\begin{eqnarray}
 \<H(n)H(0)^{\dagger}\> = \< G  \> ,
\label{deri0}
\end{eqnarray}
where $G$ is the meson propagator
\begin{equation}
G = \Tr \left[ P(\hat{\mu}_u)_{n 0}\Gamma P(\hat{\mu}_d)_{0 n}
 \Gamma^{\dagger}\right] .
\label{mesonpsop}
\end{equation}
Here $P(\hat{\mu})$ is the quark propagator at finite chemical
potential, and $\Gamma$ is the Dirac matrix which
specifies the spin of the meson.
The quark propagator is related to the Dirac operator
$D[U;\hat{\mu}]$ in the background gauge field configuration $U$ as
\begin{eqnarray}
P(\hat{\mu}) = D[U; \hat{\mu}]^{-1}.
\end{eqnarray}
The expectation value of operator $O$, $\< O \>$, stands for
\begin{eqnarray}
\< O \> = \frac{ \int [dU] O \Delta e^{-S_G} }
               { \int [dU] \Delta e^{-S_G}} ,
\label{deri3}
\end{eqnarray}
where  $S_G$ is the gluonic action and $\Delta$ is the fermion
determinant,
\begin{eqnarray}
\Delta = \det D[U; \hat{\mu}_u] \det D[U; \hat{\mu}_d] .
\label{deri2}
\end{eqnarray}
Then, the first and the second derivatives are represented as
\begin{equation}
 \frac{d}{ d\hat{\mu} } \< H(n)H(0)^{\dagger} \>
= \left\< \dot{G}+G\frac{ \dot{\Delta} }{ \Delta } \right\>
  -  \< G \>  \left\< \frac{\dot{\Delta} }{ \Delta } \right\> ,
\label{deri4}
\end{equation}
\begin{eqnarray}
 \frac{d^2}{d\hat{\mu}^2}  \< H(n)H(0)^{\dagger} \>
 &=&
   \left\< \ddot{G} + 2\dot{G}\frac{ \dot{\Delta} }{ \Delta }
           + G \frac{ \ddot{\Delta} }{ \Delta } \right\>
 \nonumber \\
 & & \hspace{-1.0cm}
  - 2  \left\< \dot{G} + G \frac{\dot{\Delta}}{\Delta} \right\>
       \left\< \frac{\dot{\Delta}}{\Delta} \right\>
 \nonumber \\
 & & \hspace{-1.0cm}
  -  \< G \>
   \left\{ \left\< \frac{\ddot{\Delta}}{\Delta} \right\>
          - 2 \left\< \frac{\dot{\Delta}}{\Delta} \right\>^2 \right\} ,
\label{deri5}
\end{eqnarray}
where the dotted $\dot{O}$ and $\ddot{O}$ denote the first and
the second derivatives of the operator $O$
with respect to $\hat{\mu}$.

At zero chemical potential, we have simpler expressions, since
\begin{equation}
 \left\< \frac{ \dot{\Delta} }{ \Delta } \right\> = 0
 \hspace{0.8cm} \mbox{for} \hspace{0.8cm}  \hat{\mu}=0  .
\label{deri6}
\end{equation}
Eq.~(\ref{deri6}) corresponds to the fact that the average baryon
number density vanishes at $\hat{\mu}=0$.
Actually, we see that
$d (\det D)/{d\hat{\mu}}= \Tr [\dot{D} D^{-1}] \det D$  is
anti-hermitian at $\hat{\mu}=0$:
\begin{eqnarray}
 \Tr \left[\dot{D}D^{-1} \right]
 &=&   \Tr \left[ \dot{D}\gamma_5\gamma_5 D^{-1}\right]
 \nonumber \\
 &=&   \Tr \left[ (-\gamma_5\dot{D^{\dagger}})
                 (D^{\dagger})^{-1}\gamma_5\right]
 \nonumber \\
 &=& - \Tr \left[ \dot{D} D^{-1} \right]^{\ast}  .
\end{eqnarray}
This means that $d(\det D)/{d\hat{\mu}}$ changes
sign under the transformation $U\rightarrow U^{\dagger}$.  Since
the measure and the gluonic action are invariant under this
transformation, its expectation value vanishes \cite{Gottlieb88}.
Thus, at zero chemical potential, Eqs.~(\ref{deri4}) and
(\ref{deri5}) turn into
\begin{eqnarray}
 \frac{d}{d\hat{\mu}}  \left\<  H(n)H(0)^{\dagger} \right\>
   &=& \left\<  \dot{G}+G \frac{\dot{\Delta}}{\Delta} \right\> ,
 \\
\frac{ d^2 }{ d\hat{\mu}^2 }  \left\<  H(n)H(0)^{\dagger} \right\>
 &=& \left\<  \ddot{G} + 2\dot{G} \frac{ \dot{\Delta} }{ \Delta }
    +G \frac{\ddot{\Delta} }{ \Delta } \right\>
 \nonumber \\
 & &  -  \< G \>  \left\< \frac{ \ddot{\Delta} }{ \Delta } \right\> .
\label{deri8}
\end{eqnarray}

We investigate derivatives with respect to both  the isoscalar and
the isovector types of chemical potential.
The isoscalar chemical potential is conjugate to the total quark
density.
In this paper we study
the response to the isoscalar chemical potential by setting
\begin{equation}
\hat{\mu}_S = \hat{\mu}_u = \hat{\mu}_d  ,
\label{scal}
\end{equation}
and for the isovector case
\begin{equation}
\hat{\mu}_V=\hat{\mu}_u=-\hat{\mu}_d  .
\label{vect}
\end{equation}
Note that Son and Stephanov proposed a model corresponding to the
isovector case as a good test bed for chemical potential
effects in QCD \cite{Son-Stephanov}.
An advantage of setting $\hat{\mu}_u=-\hat{\mu}_d$ is that the
fermion determinant is positive, and therefore
the problem becomes tractable with standard lattice techniques.
Here we do not make use of this advantage and still study the dependence
on $\hat{\mu}_V$ by performing a Taylor expansion around $\hat{\mu}_V=0$.
Future simulations at non-zero $\hat{\mu}_V$  would also constitute
a good test of the performance of our approach.

Our simulations are performed with $N_f=2$ dynamical quarks of
the staggered fermion type.
The fermion operator and its derivatives are
\begin{eqnarray}
D[U;\hat{\mu}]_{n,m}
 &=& ma\delta_{n,m}   \nonumber \\
 & & \hspace{-2.0cm}
  + \frac{1}{2} \sum_{\sigma=x,y,z} \eta_{\sigma}(n)
 \left[ U_{\sigma}(n) \delta_{n+\hat{\sigma},m}
      - U_{\sigma}^{\dagger}(n-\hat{\sigma}) \delta_{n-\hat{\sigma},m}
  \right]
 \nonumber \\
 & & \hspace{-2.0cm}
  + \frac{1}{2} \eta_{t}(n) \left[
      U_t(n) e^{\hat{\mu}} \delta_{n+\hat{t},m}
    - U_t^{\dagger}(n-\hat{t}) e^{-\hat{\mu}} \delta_{n-\hat{t},m}
    \right] , \hspace{0.8cm}
\label{KS}
\end{eqnarray}
\begin{eqnarray}
 \frac{d D}{d \hat{\mu}}
  = \frac{\eta_{t}(n)}{2}  \left[
     U_t(n) e^{\hat{\mu}} \delta_{n+\hat{t},m}
   + U_t^{\dagger}(n\!-\!\hat{t}) e^{-\hat{\mu}} \delta_{n-\hat{t},m}
   \right]  \hspace{0.2cm}
\label{KS1}
\end{eqnarray}
and
\begin{eqnarray}
 \frac{d^2 D}{d \hat{\mu}^2}
   = \frac{\eta_{t}(n)}{2}  \left[
     U_t(n)e^{\hat{\mu}}\delta_{n+\hat{t},m}
    -U_t^{\dagger}(n\!-\!\hat{t}) e^{-\hat{\mu}}\delta_{n-\hat{t},m}
    \right], \hspace{0.2cm}
\label{KS2}
\end{eqnarray}
where $\hat{\sigma}$ and $\hat{t}$ are unit vectors pointing along
spatial and temporal directions, respectively.

Taking into account the four-fold degeneracy of the
staggered fermion operator, the determinant factor $\Delta$ of $N_f=2$
fermions is then
\begin{equation}
\Delta = \exp \left( \frac{1}{4} \Tr \, \mbox{ln} D[U; \hat{\mu}_u]
     + \frac{1}{4} \Tr \, \mbox{ln} D[U; \hat{\mu}_d]
        \right) .
\label{del}
\end{equation}
Explicit formulae for the isoscalar and the isovector cases are
presented in Appendices A and B respectively.
Appendix C gives the specific forms for staggered quarks.

\section{Numerical results}
 \label{sec3}

\subsection{Lattice setup and meson screening masses}

\begin{figure}
\includegraphics[width=8.5cm]{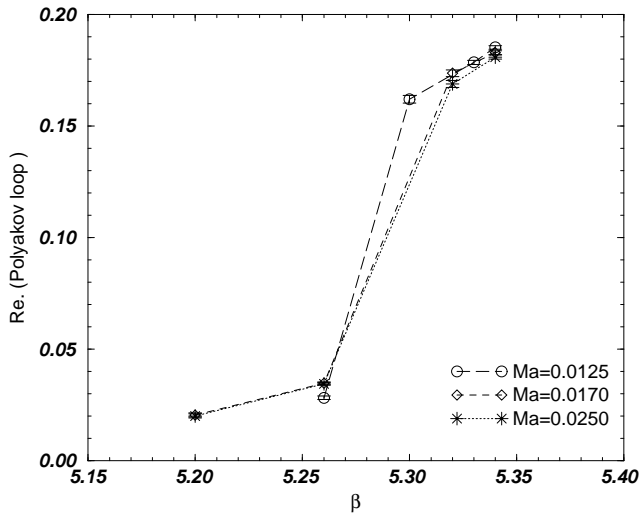}
\caption{Average value of the real part of Polyakov loop
as a function of $\beta$.}
\label{fig:pol}
\end{figure}

In this section, we describe the results of Monte Carlo (MC) simulations
for the response of meson screening masses with respect to the chemical
potentials below and above the chiral phase transition.
Numerical simulations are performed on  lattices of size
$16\times 8^2 \times 4$ with the standard Wilson gauge action and
with two dynamical flavors of staggered quarks.
The configurations are generated with the R-algorithm with  molecular
dynamics step $\delta=0.01$ and  trajectory length of 50 steps.
For $N_t=4$, the critical coupling $\beta_c$ of the finite temperature
phase transition has been determined as $\beta_c=5.271$ for $ma=0.0125$ and
$\beta_c=5.288$ for $ma=0.025$, respectively
\cite{fukugita90,gottlib87}.
We perform numerical simulations at $ma = 0.0125$, $0.017$ and
$0.025$.
The values of $\beta$ are listed in Table~\ref{tab:parameters}
together with the number of configurations and rough estimates
of the corresponding temperatures.
To evaluate the traces of the various fermionic operators, the $Z_2$ noise
method \cite{Z2} is used, with $200$ noise vectors.
The real part of the average Polyakov loop, as shown in
Figure~\ref{fig:pol}, exhibits a rapid increase between about $\beta=5.27$
and $5.28$, as expected.

\begin{table}
\caption{
Parameters of the simulations. $N_{conf}$ denotes the number of
configurations on which the correlators are measured. }
\begin{ruledtabular}
\begin{tabular}{cccc}
$ma$   & $\beta$ & $N_{conf}$ & $T/T_C$ \\
\hline
0.0125 & 5.26 & 600 & 0.99 \\
       & 5.34 & 300 & 1.10 \\
\hline
0.0170 & 5.26 & 600 & 0.98 \\
       & 5.34 & 300 & 1.09 \\
\hline
0.0250 & 5.20 & 600 & 0.89 \\
       & 5.26 & 600 & 0.96 \\
       & 5.32 & 300 & 1.05 \\
       & 5.34 & 300 & 1.07 \\
\end{tabular}
\end{ruledtabular}
\label{tab:parameters}
\end{table}

\begin{figure}
\includegraphics[width=8.5cm]{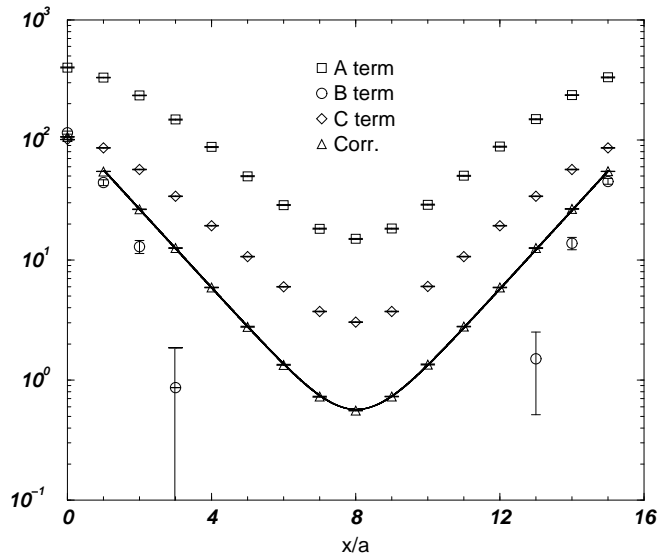}
\caption{
Pseudoscalar correlator and several contributions to its
responses to $\hat{\mu}_{S,V}$ (see Eq.~(\ref{ABD}))
at $\beta=5.34$ and $ma=0.025$. The correlator is fitted to the single
pole form, Eq.~(\ref{singlepole}).}
\label{2ndps534025}
\end{figure}

\begin{figure}
\includegraphics[width=8.5cm]{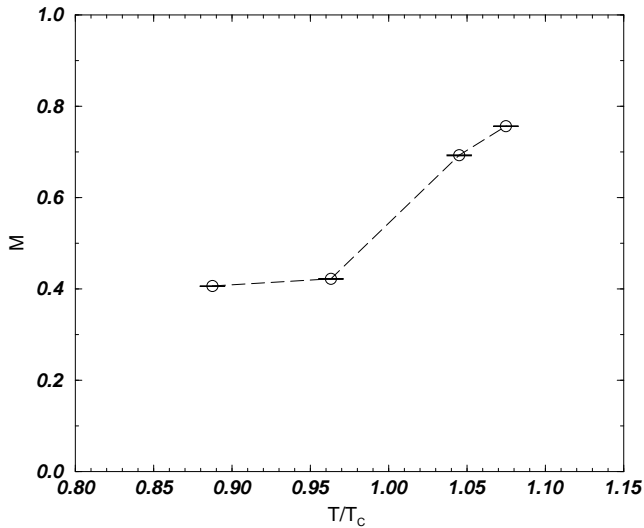}
\caption{
Screening mass of the PS meson in lattice units at $ma=0.025$.
$\beta_C$ is taken from \cite{gottlib87}.
Values of $T/T_C$ are estimated by using the two-loop $\beta$-function. }
\label{mass025}
\end{figure}

Meson correlators are measured on configurations separated by
$20$ trajectories.
We use a corner-type wall source \cite{Marinari} after Coulomb gauge
fixing in each $(y,z,t)$-hyperplane.
We determine the meson screening masses $\hat{M}$ by fitting
the MC data to the single exponential form, Eq.~(\ref{singlepole}),
with the fitting range $x=1$--$15$.
As shown in Figure~\ref{2ndps534025}, the fit is successfully applied
for the pseudoscalar(PS) meson channel.
The extracted PS
meson screening mass is listed in Table~\ref{tab5h}.
Below $T_c$, the present lattice extent in the shorter spatial directions
is not sufficiently large to  extract reliably  the PS meson mass.
Since it is known that the finite size effects on hadron masses are
rather severe with dynamical staggered fermions \cite{FSE1,FSE2},
one should use lattices with $L\cdot m_{\pi}$ larger than $\sim 5$
to avoid finite size effects, and hence larger lattice sizes are
obviously required for a quantitative analysis.
Goals of the present study are, however, to show the applicability of
the method to extract the responses of meson mass to the
chemical potentials, and to observe the qualitatively different behavior
below and above $T_c$ giving us hints for future detailed studies.
Therefore we present also the results below $T_c$
in spite of the large systematic uncertainty.
The screening mass of the PS meson at $ma =
0.025$ as a function of $T/T_c$ is shown in Figure~\ref{mass025}.

\begin{table}
\caption{
Responses of the PS meson to the isoscalar chemical
potential $\hat{\mu}_S$. }
\begin{ruledtabular}
\begin{tabular}{cccccc}
$ma$ &  $\beta$ &  $\hat{M}$  &
 $\frac{1}{A}\frac{d^2A}{d\hat{\mu}^2}$ &
 $\frac{d^2\hat{M}}{d\hat{\mu}^2}$      &
 $\frac{1}{\hat{\gamma}}\frac{d^2\hat{\gamma}}{d\hat{\mu}^2}$ \\
\hline
0.0125 & 5.26 & 0.2956( 2) & -1.4(20) & 0.17(35)& -0.8(23) \\
       & 5.34 & 0.7513(11) & -4.23(49)& 5.39(10)&  2.95(51)\\
\hline
0.0170 & 5.26 & 0.3506( 2) & -1.5(14) & 0.30(26)& -0.6(16) \\
       & 5.34 & 0.7421(35) & -3.68(75)& 5.82(16)&  4.16(78)\\
\hline
0.0250 & 5.20 & 0.4061( 2) & -0.4(14) & 0.16(26)&  0.0(16) \\
       & 5.26 & 0.4218( 2) & -0.8(11) & 0.29(20)& -0.1(12) \\
       & 5.32 & 0.6926(11) & -4.65(91)& 5.17(20)&  2.82(96)\\
       & 5.34 & 0.7534( 7) & -3.17(41)& 4.43( 8)&  2.71(42)\\
\end{tabular}
\end{ruledtabular}
\label{tab5h}
\end{table}

In the case of the vector meson correlator, although the procedure
is in principle also applicable,
the statistical fluctuations
prevent us from reaching any relevant conclusion at this stage,
and therefore we do not report results for the vector
channel in the following.

\subsection{Responses of the PS meson mass to the isoscalar
chemical potential}

Several terms contribute to the derivatives of the meson correlator
with respect to the chemical potential: see Eqs.~(\ref{vec1st}),
(\ref{sca2nd2}), and (\ref{vec2nd2e}) in Appendices.
As representative terms, in Figure~\ref{2ndps534025} we show
\begin{eqnarray}
(A): & &
   2 \sum_{y,z,t} \Re \left\< \Tr [ (P\dot{D}P)_{n 0}
                      (P\dot{D}P)^{\dagger}_{n 0} ] \right\> ,
 \nonumber \\
(B): & &
   4 \sum_{y,z,t} \Re \left\< \Tr [(P\dot{D}P\dot{D}P)_{n 0}
                            P_{n 0}^{\dagger} ] \right\> ,
 \nonumber \\
(C): & &
   2 \sum_{y,z,t} \Re \left\< \Tr [ (P\ddot{D}P)_{n 0}
                            P_{n 0}^{\dagger} ] \right\> .
\label{ABD}
\end{eqnarray}
All of them are determined with reasonable accuracy.

\begin{figure}
\includegraphics[width=8.5cm]{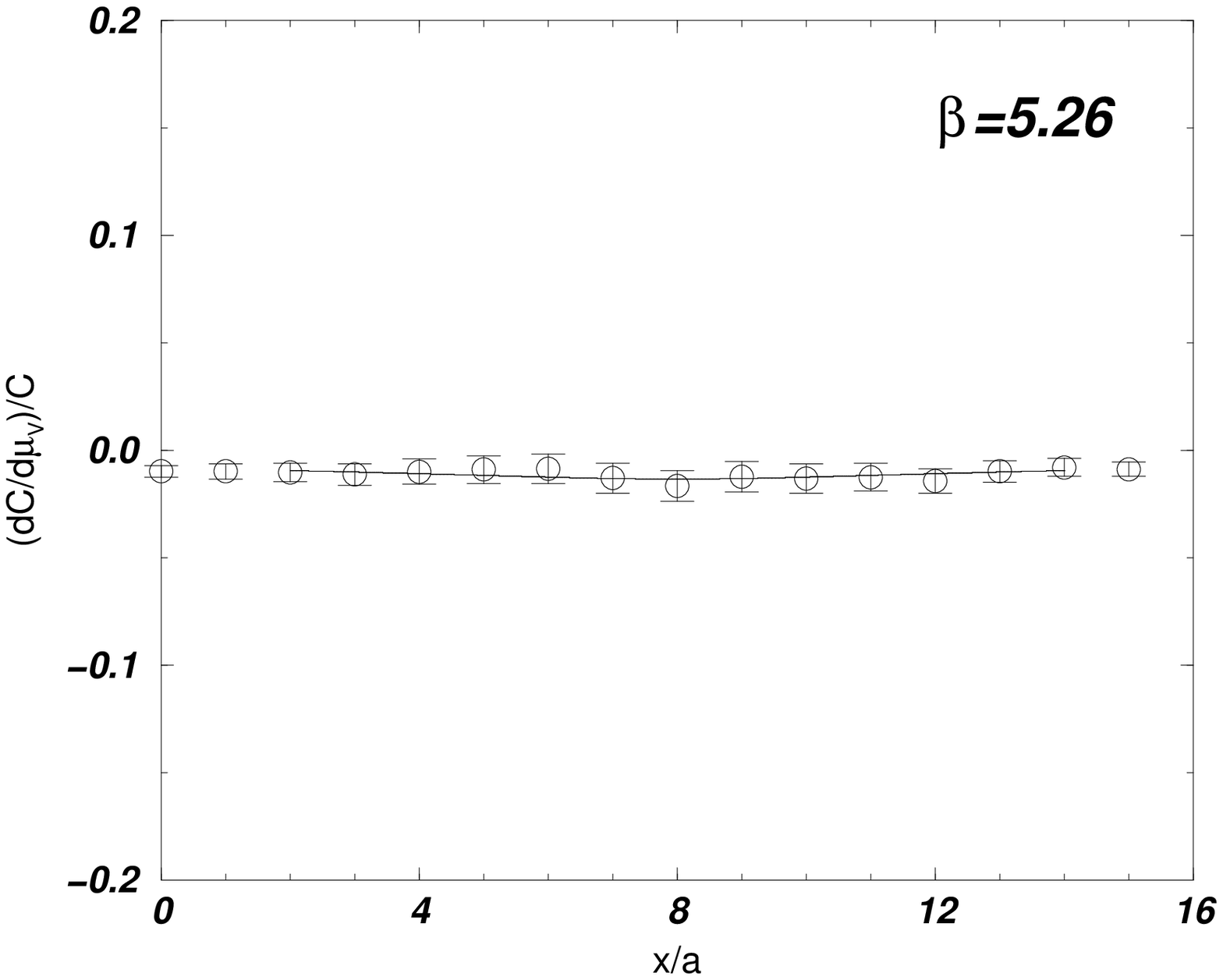}
\includegraphics[width=8.5cm]{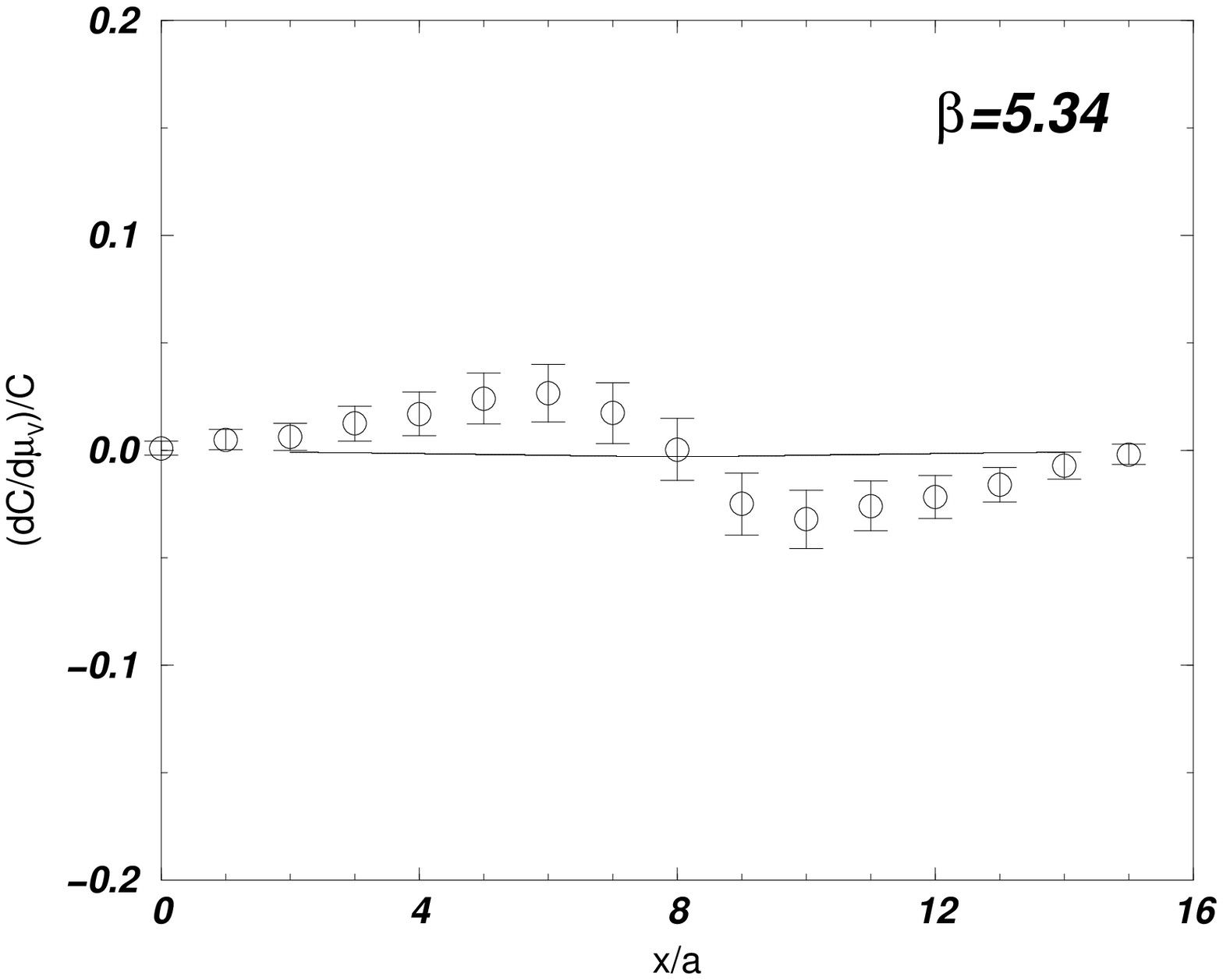}
\caption{
The first response of the PS meson correlator at
$\beta=5.26$ ($T<T_c$, top figure) and $\beta=5.34$ ($T>T_c$, bottom).
The quark
mass is $ma=0.025$. }
\label{1stps025}
\end{figure}

Let us start with the first response of the PS correlator to the
chemical potential. Note that the first derivative with respect to
the isoscalar chemical potential is identically zero (see
Eq.~(\ref{sca1st}) in Appendix A). For the isovector chemical
potential, Figure~\ref{1stps025} shows $C^{-1} d C / d \hat{\mu}_V
$  at $\beta=5.26$ and $5.34$. At both temperatures, the values
are very small. This is consistent with a comparative study in the
Nambu--Jona-Lasinio model, which also predicts very small
responses around the critical temperature \cite{CHOE}.

\begin{figure}
\includegraphics[width=8.5cm]{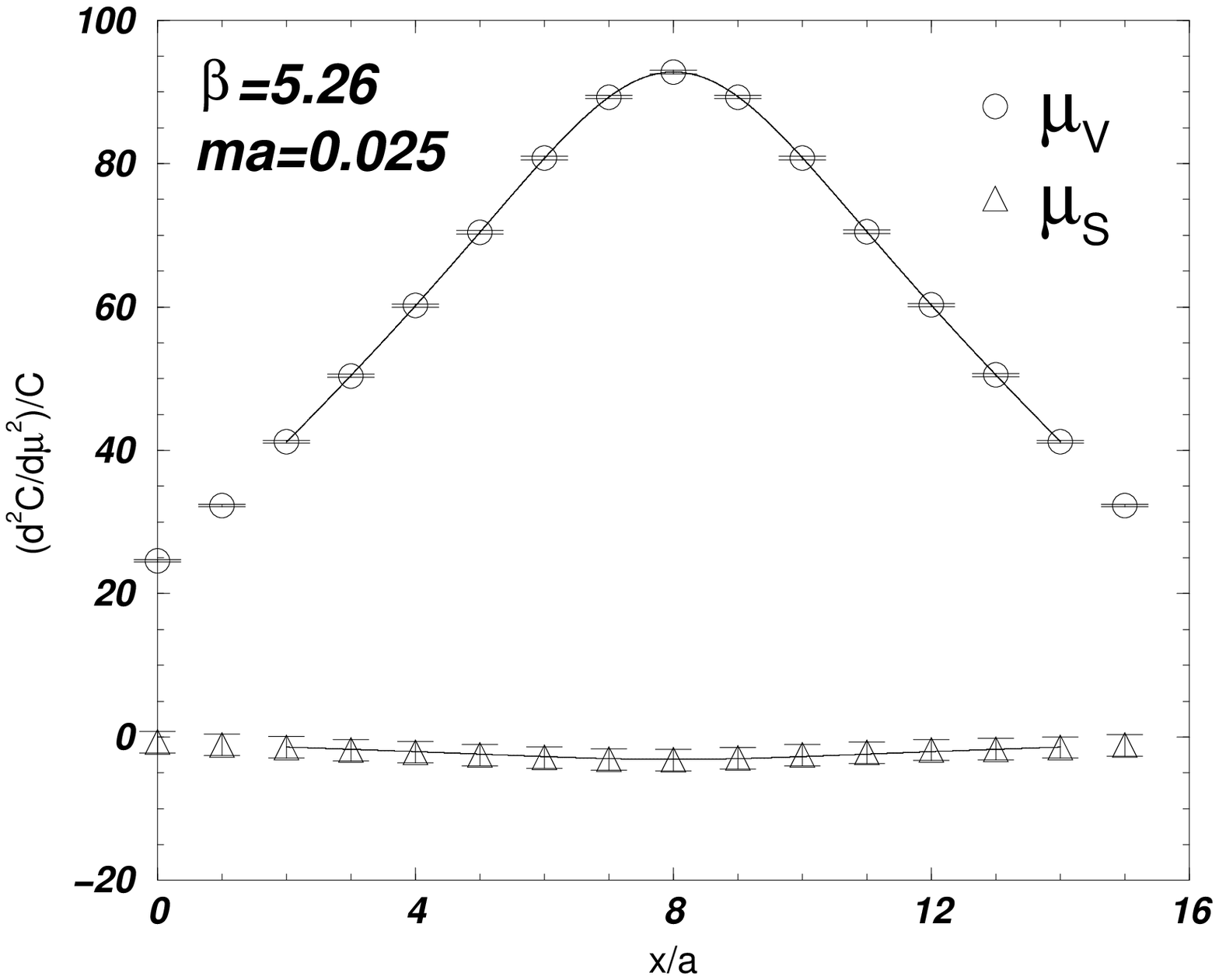}
\includegraphics[width=8.5cm]{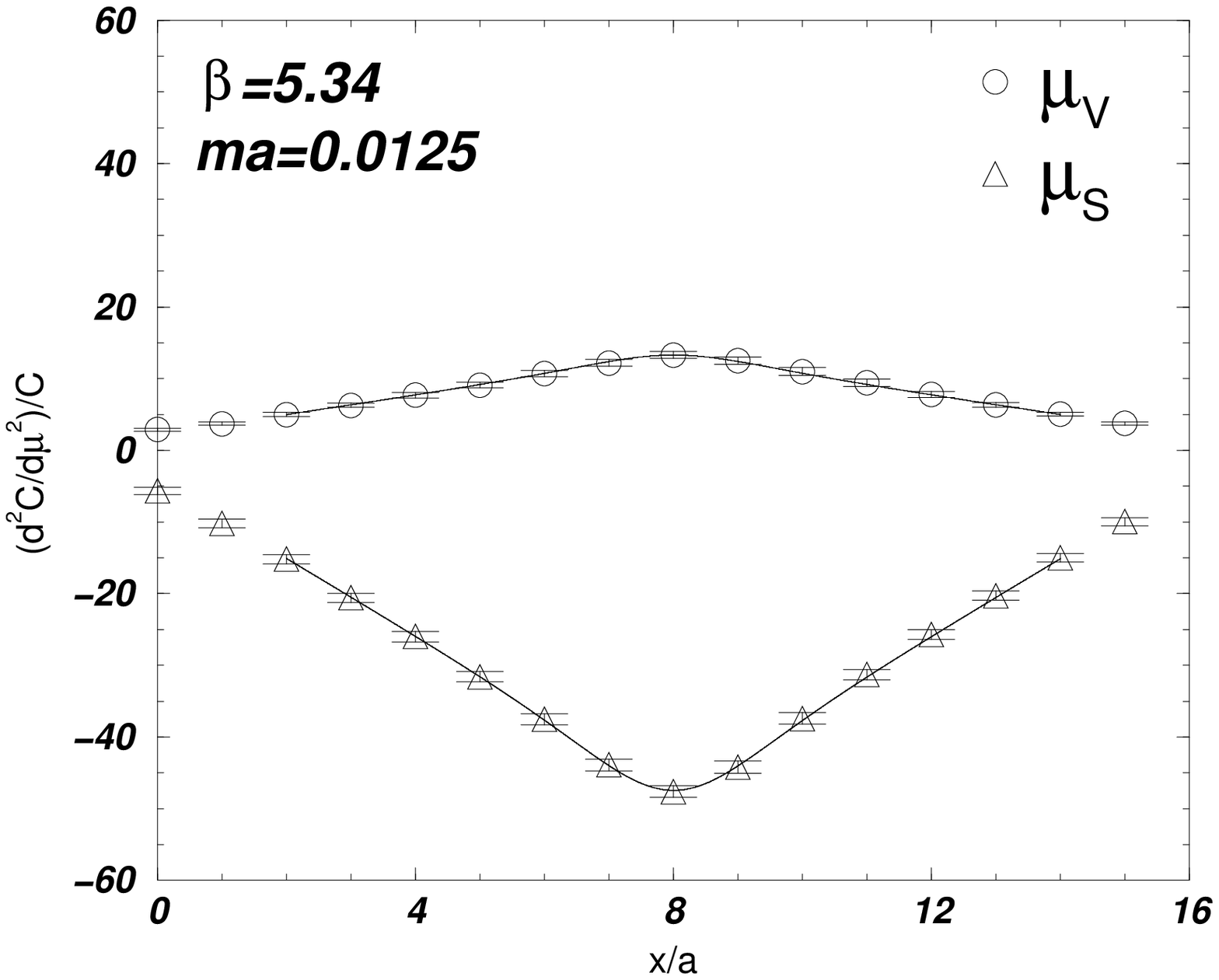}
\caption{
The second responses of the PS
meson correlator at $\beta=5.26$ and $ma=0.025$ ($T<T_c$, top), and
at $\beta=5.34$ and $ma=0.0125$ ($T>T_c$, bottom).
The curves are fits to Eq.~(\ref{deri-2}). }
 \label{2ndps526025b}
\end{figure}

We then consider the second responses.
Figure~\ref{2ndps526025b} shows $C^{-1} d^2 C / d \hat{\mu}^2 $
at $\beta=5.26$ (below $T_c$) and $5.34$ (above $T_c$) for the isoscalar
and isovector chemical potentials.
The solid curves represent the fits to Eq.~(\ref{deri-2}),
after fitting $C(x)$ to Eq.~(\ref{singlepole}).

We determine the responses of the meson mass with respect to the chemical
potentials as follows.
As described in the last subsection,
the meson correlator is first fitted to Eq.~(\ref{singlepole}) to extract the
screening mass, $\hat{M}$.
Then, we fit the MC results for the derivatives of the meson correlator
to Eqs.~(\ref{deri-1}) and (\ref{deri-2}), substituting the
determined value of $\hat{M}$.
The derivatives of mass and coupling are then obtained as fitting
parameters.
Note that for $\hat{\mu}_S$ we omit the fitting step to
Eq.~(\ref{deri-1}), since the first response is strictly zero.

\begin{figure}
\includegraphics[width=8.5cm]{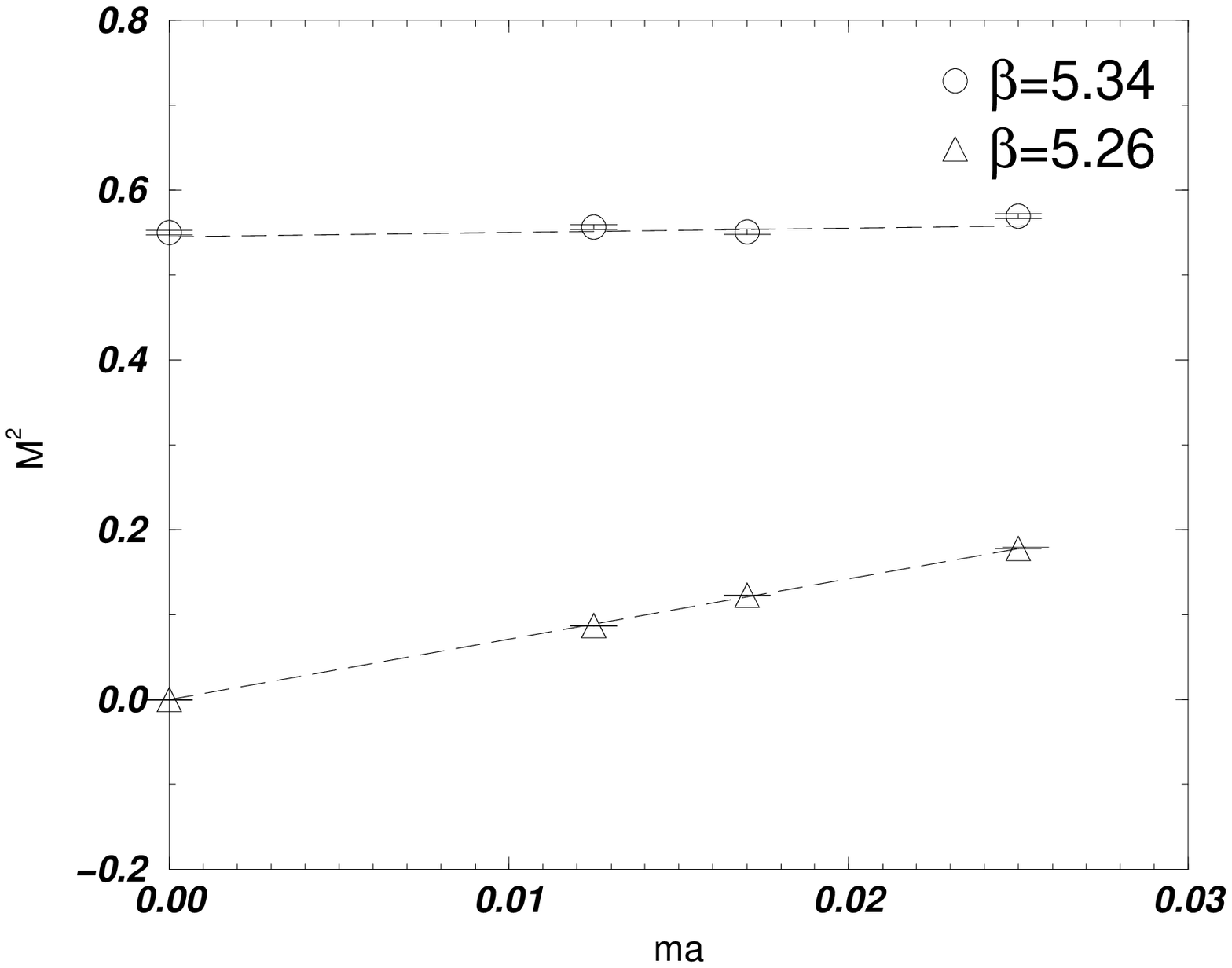}
\includegraphics[width=8.5cm]{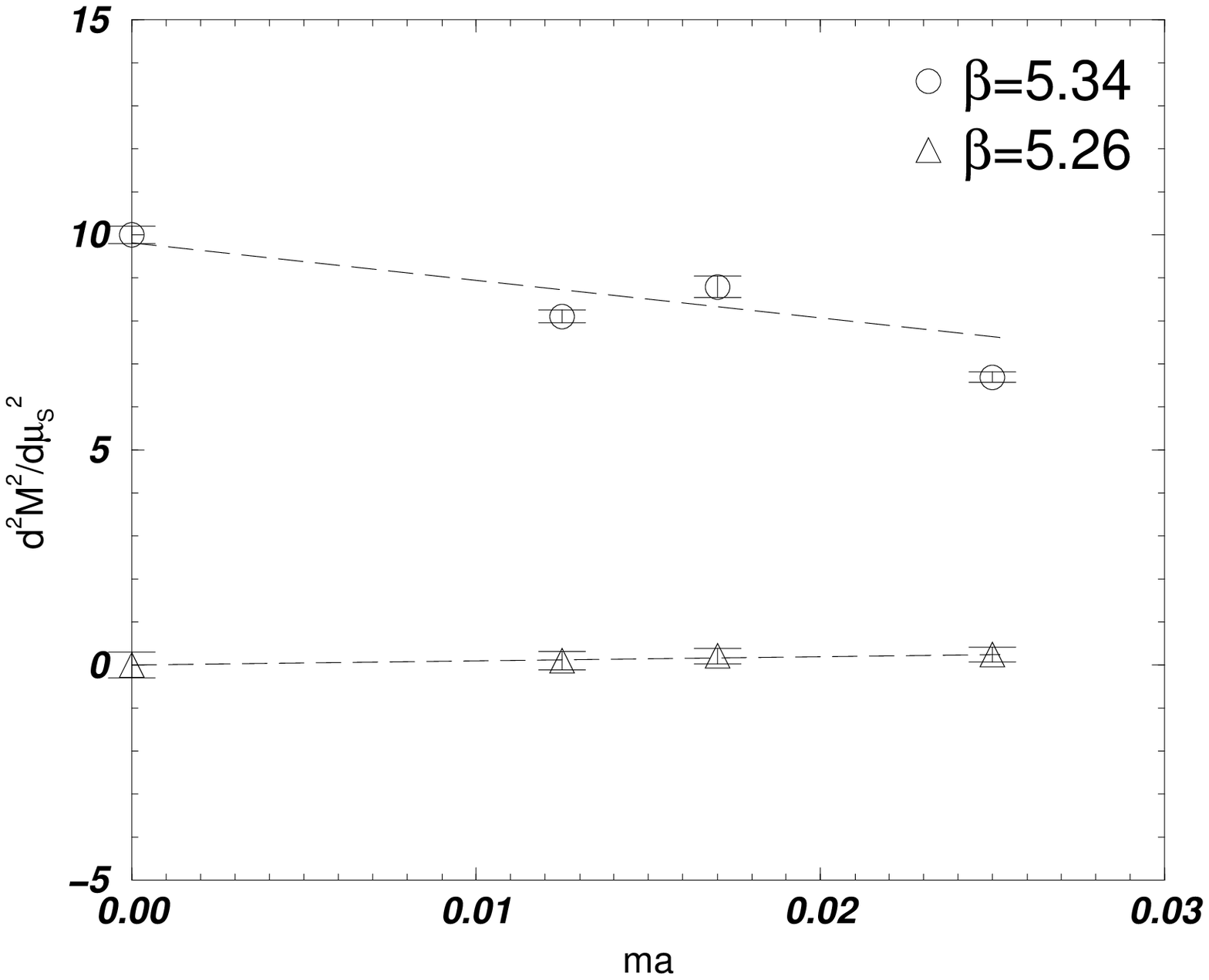}
\caption{
$\hat{M}^2$ (top) and $d^2\hat{M}^2/d\hat{\mu}_S^2$ (bottom) for
the PS meson versus $ma$.
$\beta$ is $5.26$ (triangles) and $5.34$ (circles).
Extrapolation to $ma=0$ is also shown.}
\label{chlimddotM}
\end{figure}

Results of the PS meson response
to the isoscalar chemical potential are summarized in Table
\ref{tab5h}.
In the low temperature phase, the dependence of the mass on $\hat \mu_S$
is small.
This behavior is to be expected, since, below the critical
temperature and in the vicinity of zero $\hat \mu_S$, the PS
meson is still a Goldstone boson.
In fact, the extrapolated value to the chiral limit of the isoscalar
response is consistent with zero, as shown in Figure~\ref{chlimddotM}.
This is in contrast with the behavior above $T_c$, where
$d^2 \hat{M}^2/d \hat{\mu}^2$ seems to remain finite
even in the chiral limit.
In addition, our results suggest that the response of the coupling
is small below $T_c$.

Above $T_c$, we first note that the correlator and its response
are still well fitted by the single pole formulae,
Eqs.~(\ref{singlepole}--\ref{deri-2}).
The screening masses are manifestly larger than those below $T_c$.
This is consistent with previous work \cite{Gott}.
As pointed out above, the response of the mass above $T_c$ becomes large,
as a reflection of the fact that the pion is no longer a Goldstone
boson and as an indication of chiral symmetry restoration.
We also find that the response of the coupling $\hat{\gamma}$ increases
above $T_c$, as shown in Figure~\ref{residue025}, which may indicate
a larger overlap of the wall-source operator with free quarks
in the plasma phase.
Note that $(d^2 \hat{\gamma} /d \hat{\mu}^2)/ \hat{\gamma}$
does not depend on the choice of the source normalization.

\begin{figure}
\includegraphics[width=8.5cm]{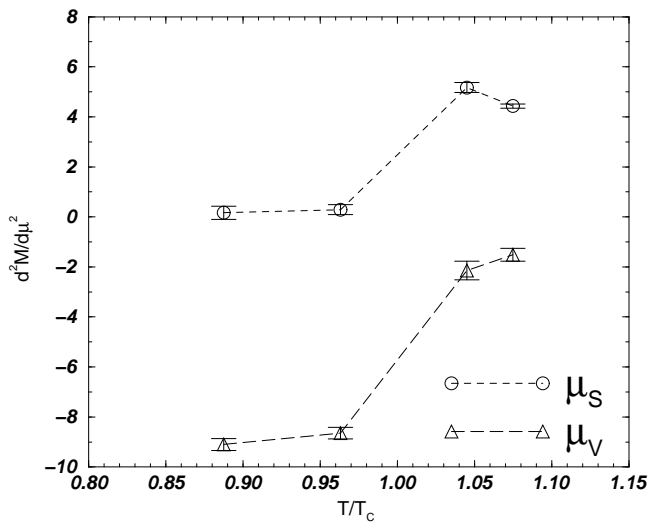}
\caption{Second responses $d^2\hat{M}/d\hat{\mu}^2_S$ and
$d^2\hat{M}/d\hat{\mu}^2_V$ of the PS meson mass at $ma=0.025$.
$T/T_c$ is estimated as in Figure~\ref{mass025}.}
\label{fig:ddotM025}
\end{figure}

\subsection{Responses of the PS meson mass
            to the isovector chemical potential}

\begin{figure}
\includegraphics[width=8.5cm]{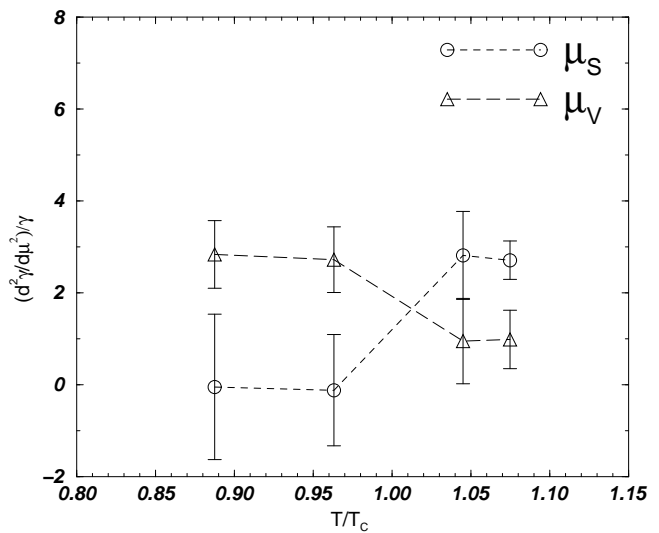}
\caption{Second responses of the couplings,
$\hat{\gamma}^{-1}d^2\hat{\gamma}/d\hat{\mu}^2_S$,
$\hat{\gamma}^{-1}d^2\hat{\gamma}/d\hat{\mu}^2_V$, for the
PS meson at $ma=0.025$.
$T/T_c$ is estimated as in Figure~\ref{mass025}.}
\label{residue025}
\end{figure}

Results for the isovector chemical potential are summarized in
Table \ref{tab6h}.
In the presence of the isovector chemical potential, $\pi^{+}$ and
$\pi^{-}$ may have different masses. Here we consider the $\pi^{+}$
 ( $u \bar{d}$ ) meson  as shown in Eq.~(\ref{mesonpsop}).
In contrast to the case of the isoscalar
chemical potential, the second order response of the mass is
significantly large in the low temperature phase, and decreases in magnitude
above $T_c$. The difference between the isoscalar and isovector chemical
potentials is illustrated in Figure~\ref{fig:ddotM025}.
The response of  $\hat{\gamma}$ also shows different behaviors in the
confined and deconfined phases, illustrated in Figure~\ref{residue025}.

These features are manifest even for a small quark mass parameter.
Note that the isovector potential explicitly breaks the $u$-$d$
symmetry, even if the two quarks have equal masses.
The phase structure in the $(T,| \mu_V|)$
plane has been studied by Son and Stephanov \cite{Son-Stephanov}.
The original $SU(2)_{L+R}$ symmetry at non-zero
quark mass and zero chemical potential is broken down to $U(1)_{L+R}$.
At zero $T$ and  for $| \mu_V|$ larger than the mass
of the pseudoscalar, the system is in a
different phase than at $ \mu_V=0$. The ground state is a pion
condensate and there is one massless Goldstone boson associated with the
spontaneous breaking of the $U(1)_{L+R}$ symmetry.
For $| \mu_V|= m_{PS}$, the critical temperature is $T=0$.
At sufficiently high temperature, the condensate melts and the
symmetry is restored.
Due to the presence of the phase boundary, we do not expect to be able
to reach the condensed phase by Taylor expanding around $ \mu_V=0$.
We can, however, hope to get some hints about the presence of
the phase boundary while keeping $| \mu_V|<m_{PS}$.
In this case, the system is in the same ground
state as for zero chemical potential, there are no exact Goldstone modes
and the three pions are massive.

An interesting point in this respect is that the second derivative of the mass
is negative in the low temperature phase, in marked contrast with
what happened for the isoscalar potential. The mass tends to
decrease under the influence of the isovector chemical potential,
reflecting the fact that for low temperature and chemical potential
above the pion mass,
a Goldstone mode appears \cite{Stephanov,Son-Stephanov}.
This is more clearly shown by an expansion as in
Eq.~(\ref{exp}).  At $\beta=5.26$ and $ma=0.017$, the
data suggest
\begin{eqnarray}
 \left. \frac{M(\mu_V)}{T}\right|_{\mu_V}
 &=& ( 1.4024 \pm 0.0008 )  \nonumber\\
 & &  \hspace{-1.0cm}
    + ( -0.0005 \pm 0.0010) \left( \frac{\mu_V}{T} \right)
 \nonumber \\
 & &  \hspace{-1.0cm}
   -( 1.31 \pm 0.04 ) \left( \frac{\mu_V}{T} \right)^2
   + O \left[ \left(\frac{\mu_V}{T}\right)^3 \right].
  \hspace{1.0cm}
\end{eqnarray}
The coefficient of the linear term is consistent with zero.
Notice also in Table III that the lighter the quark mass, the stronger
the response, a possible indication that for lighter pions the phase boundary
is closer to the zero chemical potential axis, as suggested in
\cite{Son-Stephanov}.

In the high temperature phase, the dependence of the masses on $\mu_V$
decreases. Since the PS meson becomes heavier, the phase boundary
to the pion condensate phase is farther away from the $ \mu_V=0$ axis.
The weaker responses may be understood from this point of view.

\begin{table*}
\caption{Responses of the PS meson correlator to the isovector
chemical  potential $\hat{\mu}_V$.}
\begin{ruledtabular}
\begin{tabular}{ccccccc}
$ma$ & $\beta$&$\frac{1}{A}\frac{dA}{d\hat{\mu}}$ &
$\frac{d\hat{M}}{d\hat{\mu}}$ &
$\frac{1}{A}\frac{d^2A}{d\hat{\mu}^2}$&
$\frac{d^2\hat{M}}{d\hat{\mu}^2}$&
$\frac{1}{\hat{\gamma}}\frac{d^2\hat{\gamma}}{d\hat{\mu}^2}$\\
\hline
0.0125& 5.26&  ~0.0029(57)&$-$0.0001(12)& 47.46(71)&$-$12.93(43)& 3.7(16)~\\
      & 5.34&  ~0.0047(93)&  ~0.0006(21)& ~2.32(64)&~$-$1.32(32)& 0.56(77)\\
\hline
0.0170& 5.26&$-$0.0081(48)&$-$0.0005(10)& 33.52(61)&$-$10.49(33)& 3.6(11)~\\
      & 5.34&  ~0.0000(82)&$-$0.0012(19)& ~2.74(64)&~$-$1.48(32)& 0.75(78)\\
\hline
0.0250& 5.20&  ~0.0062(39)&  ~0.0016(~8)& 25.24(46)&~$-$9.10(23)& 2.84(74)\\
      & 5.26&$-$0.0080(37)&  ~0.0007(~8)& 23.22(46)&~$-$8.64(23)& 2.72(71)\\
      & 5.32&$-$0.0054(64)&$-$0.0020(14)& ~4.04(75)&~$-$2.14(38)& 0.95(93)\\
      & 5.34&  ~0.000(6)~~&  ~0.000(1)~~& ~2.99(53)&~$-$1.51(26)& 0.99(60)\\
\end{tabular}
\end{ruledtabular}
\label{tab6h}
\end{table*}

\section{Conclusions}
 \label{sec4}

In this work, we have developed a framework to study the response
of hadrons to the chemical potential.
It is based on Taylor expanding hadronic quantities about $\mu=0$.
We show the first results of the first and second derivatives of
the PS meson screening mass with respect to $\mu$.
As shown in the previous section, the second order responses are
sizable and reveal several characteristic features.
The behavior of the responses for the PS meson seems to have
close contact to the chiral symmetry restoration.
For the isoscalar chemical potential $\mu_S$, the dependence
of the PS mass on $ \mu_S$  in the chiral limit is
consistent with zero, reflecting the fact that at low temperature and
small $ \mu_S$ the pion  is still a Goldstone boson.
For the isovector chemical potential, we show features consistent with
the phase structure proposed by Son and Stephanov
\cite{Son-Stephanov}.
The $u \bar{d}$ PS meson mass tends to decrease as a function
of $ \mu_V$ at a much stronger rate in the low temperature phase.

It is notable that a
single hadron pole gives a good description for the response as
well as for the correlator even at $\beta=5.34$ ($T/T_c \approx 1.1$)
in the PS channel.

Since the present study is a first trial, our simulations have
been performed on a rather small lattice.
Since the difference between the dynamics of $N_t=4$ and $N_t=6$
lattices is sizable \cite{cppacs01}
and since the spatial volume should be large enough to accommodate the hadrons,
further investigations on larger lattices are indispensable.
The chemical potential responses of other quantities, such as the
nucleon mass and the quark condensate, are important issues.
Exploratory studies on these subjects are in progress.

\begin{acknowledgments}

This work is supported by Grant-in-Aide for Scientific Research by
Monbusho, Japan (No.11694085, No.11740159, and No. 12554008).
Simulations were performed on the HITACHI SR8000 at IMC, Hiroshima
University.
H.~M. is supported by the Japan Society for the Promotion of Science
for Young Scientists.

\end{acknowledgments}

\appendix

\section{Formulae for the isoscalar chemical potential response}

The quark propagator $P$ satisfies the relation
\begin{eqnarray}
 P(\hat{\mu}_d)_{0 n}=\gamma_5
 P^{\dagger}(-\hat{\mu}_d)_{n 0}\gamma_5 ,
\label{gamma5}
\end{eqnarray}
so that the meson correlator $G$ is given by
\begin{eqnarray}
 G = \Tr \left[ P(\hat{\mu}_S)_{n 0}\Gamma \gamma_5
  P(-\hat{\mu}_S)_{n 0}^{\dagger} \gamma_5 \Gamma^{\dagger} \right] ,
\end{eqnarray}
where ``$\Tr$'' means the trace over spinor and color indices.
Each propagator is expanded as
\begin{eqnarray}
P(\hat{\mu})  &=&  P - \hat{\mu} P\dot{D}P +
  \frac{\hat{\mu}^2}{2} (2 P \dot{D}P\dot{D}P - P\ddot{D}P )
 \nonumber \\
  & &  + O(\hat{\mu}^3) , \\
P(-\hat{\mu}) &=&  P + \hat{\mu} P\dot{D}P +
  \frac{\hat{\mu}^2}{2} (2P\dot{D}P\dot{D}P - P\ddot{D}P )
 \nonumber \\
  & &  + O(\hat{\mu}^3) ,
  \hspace{0.4cm}
\end{eqnarray}
where $P$ and $D$ are the quark propagator and the Dirac operator at
zero chemical potential, respectively, and the relation
\begin{equation}
\dot{P} = - P \dot{D} P
\end{equation}
is used.

The first derivative at $\hat{\mu}_S=0$ is
\begin{equation}
\dot{G}  =  -2 i \Im \Tr \left[ (P\dot{D}P)_{n 0} \Gamma
    \gamma_5 P_{n 0}^{\dagger} \gamma_5\Gamma^{\dagger} \right] ,
\label{deri9}
\end{equation}
and the second derivative at $\hat{\mu}_S=0$ is obtained as
\begin{eqnarray}
\ddot{G} &=& \   4 \Re \Tr \left[ (P\dot{D}P\dot{D}P)_{n 0}\Gamma
           \gamma_5 P_{n 0}^{\dagger} \gamma_5\Gamma^{\dagger} \right]
 \nonumber  \\
 & & - 2 \Re \Tr \left[  (P\ddot{D}P)_{n 0}\Gamma
           \gamma_5 P_{n 0}^{\dagger} \gamma_5 \Gamma^{\dagger} \right]
 \nonumber  \\
 & & - 2 \Tr \left[ (P\dot{D}P)_{n 0} \Gamma \gamma_5
   (P\dot{D}P)_{n 0}^{\dagger} \gamma_5\Gamma^{\dagger} \right] .
\label{deri11}
\end{eqnarray}

Let us calculate the derivatives of the fermionic determinant
$\Delta$.
Using the following equations,
\begin{eqnarray}
\frac{d}{d\hat{\mu}} \det D  &=& \Tr [ \dot{D}P ] \det D ,
 \nonumber \\
\frac{d^2}{d\hat{\mu}^2} \det D  &=&
  \left\{ \Tr [ \ddot{D} P ] - \Tr [ \dot{D} P \dot{D} P ]
        + \Tr [ \dot{D} P ]^2 \right\}
 \nonumber \\
 & & \times  \det D ,
 \label{deri12i}
\end{eqnarray}
we have
\begin{eqnarray}
\frac{ \dot{\Delta} }{ \Delta }
  &=&  2 \Tr \left[ \dot{D} P         \right] ,
 \nonumber \\
\frac{ \ddot{\Delta} }{ \Delta }
  &=&  2 \Tr \left[ \ddot{D}P        \right]
      -2 \Tr \left[ \dot{D}P\dot{D}P \right]
      +4 \Tr \left[ \dot{D}P         \right]^2 .
\label{deri12}
\end{eqnarray}

Combining Eqs.~(\ref{deri8}), (\ref{deri9}), (\ref{deri11}),
(\ref{deri12i}) and (\ref{deri12}), we have
\begin{eqnarray}
\frac{d}{d\hat{\mu}} \Re \< H(n)H(0)^{\dagger} \> = 0,
\label{sca1st}
\end{eqnarray}
and
\begin{widetext}
\begin{eqnarray}
 \frac{d^2}{d\hat{\mu}^2} \Re \< H(n)H(0)^{\dagger} \>
 &=&
   4 \Re \left\< \Tr \left[ (P\dot{D}P\dot{D}P)_{n 0} \Gamma \gamma_5
         P_{n 0}^{\dagger} \gamma_5 \Gamma^{\dagger} \right] \right\>
 - 2 \Re \left\< \Tr \left[ (P\ddot{D}P)_{n 0}\Gamma
      \gamma_5 P_{n 0}^{\dagger} \gamma_5\Gamma^{\dagger}
                                                     \right] \right\>
\nonumber \\
 & &  \hspace{-3cm}
 - 2 \Re \left\< \Tr \left[ (P\dot{D}P)_{n 0}\Gamma \gamma_5
    (P\dot{D}P)_{n 0}^{\dagger} \gamma_5\Gamma^{\dagger}
                                                     \right] \right\>
 + 8 \left\<
          \Im \Tr \left[ (P\dot{D}P)_{n 0}\Gamma \gamma_5
               P_{n 0}^{\dagger} \gamma_5\Gamma^{\dagger} \right]
          \Im \Tr \left[ \dot{D} P \right]                   \right\>
 \nonumber \\
 & &  \hspace{-3cm}
 + 2 \Re \left\{ \;
    \left\< \Tr \left[ P_{n 0}\Gamma \gamma_5
             P_{n 0}^{\dagger} \gamma_5\Gamma^{\dagger} \right]
        \left( \Tr [ \ddot{D}P ] - \Tr [ \dot{D}P\dot{D}P ]
        +2 \Tr [\dot{D}P]^2  \right)     \right\>  \right.
 \nonumber \\
 & & \hspace{-1.7cm}
  \left.
     - \left\< \Tr \left[ P_{n 0}\Gamma \gamma_5 P_{n 0}^{\dagger}
           \gamma_5\Gamma^{\dagger} \right] \right\>
     \left\<  \Tr [ \ddot{D}P ] - \Tr [ \dot{D}P\dot{D}P ]
                +2 \Tr [ \dot{D}P ]^2 \right\>  \; \right\} .
\label{sca2nd}
\end{eqnarray}
\end{widetext}

\section{Formulae for the isovector chemical potential response}

Next, we consider responses to the isovector chemical potential,
Eq.~(\ref{vect}). In this case, the first
derivative of $\Delta$ vanishes,
\begin{equation}
\dot{\Delta} = \left.  \frac{d}{d\hat{\mu}_V}
  \left( \det D[U; \hat{\mu}_V] \, \det D[U; -\hat{\mu}_V] \right)
  \right|_{\hat{\mu}_V=0}= 0 ,
\end{equation}
and the second derivative is obtained as
\begin{eqnarray}
\frac{ \ddot{\Delta} }{ \Delta } =
   2 \Tr [\ddot{D}P] - 2 \Tr [ \dot{D} P \dot{D} P ] .
\label{deri15}
\end{eqnarray}

Similarly, derivatives of $G$ are calculated as
\begin{eqnarray}
 \dot{G} = - 2 \Re \Tr \left[ (P\dot{D}P)_{n 0} \Gamma
        \gamma_5 P_{n 0}^{\dagger} \gamma_5 \Gamma^{\dagger} \right] ,
\label{deri16}
\end{eqnarray}
and
\begin{eqnarray}
\ddot{G}
 &=&  4 \Re \Tr \left[ (P\dot{D}P\dot{D}P)_{n 0} \Gamma
      \gamma_5 P_{n 0}^{\dagger} \gamma_5 \Gamma^{\dagger} \right]
 \nonumber \\
 & & -2 \Re \Tr \left[ (P\ddot{D}P)_{n 0} \Gamma \gamma_5
       P_{n 0}^{\dagger} \gamma_5 \Gamma^{\dagger} \right]
 \nonumber \\
 & & +2 \Tr \left[ (P\dot{D}P)_{n 0} \Gamma \gamma_5
       (P\dot{D}P)_{n 0}^{\dagger} \gamma_5 \Gamma^{\dagger} \right] .
 \label{deri17}
\end{eqnarray}

Resulting expressions for the first and second responses to the
isovector chemical potentials are
\begin{eqnarray}
\frac{d}{d\hat{\mu}} \Re \left\< H(n)H(0)^{\dagger} \right\> & &
 \nonumber \\
 & & \hspace{-3cm}
  = -2 \Re \Tr \left[ (P\dot{D}P)_{n 0} \Gamma \gamma_5 P_{n 0}^{\dagger}
        \gamma_5\Gamma^{\dagger} \right] ,
\label{vec1st}
\end{eqnarray}
and
\begin{eqnarray}
\frac{d^2}{d\hat{\mu}^2} \Re \left\< H(n)H(0)^{\dagger} \right\> =
    \hspace{-3.8cm} & &
  \nonumber \\
 & &
  4 \Re \left\< \Tr \left[ (P\dot{D}P\dot{D}P)_{n 0}\Gamma \gamma_5
           P_{n 0}^{\dagger} \gamma_5\Gamma^{\dagger} \right] \right\>
 \nonumber \\
 & &
 -2 \Re \left\< \Tr \left[ (P\ddot{D}P)_{n 0} \Gamma \gamma_5
           P_{n 0}^{\dagger} \gamma_5\Gamma^{\dagger} \right] \right\>
\nonumber \\
 & &
 +2 \Re \left\< \Tr \left[ (P\dot{D}P)_{n 0} \Gamma \gamma_5
           (P\dot{D}P)_{n 0}^{\dagger} \gamma_5 \Gamma^{\dagger}
                                                       \right]\right\>
\nonumber \\
 & &
 +2 \Re \left\{
      \left\< \Tr \left[ P_{n 0}\Gamma \gamma_5 P_{n 0}^{\dagger}
              \gamma_5\Gamma^{\dagger} \right]
       \! \left( \Tr [ \ddot{D}P ] - \Tr [ \dot{D}P\dot{D}P ] \right)
                                                   \right\> \right.
\nonumber \\
 & & \  \left.
  - \left\<  \Tr \left[ P_{n 0} \Gamma \gamma_5 P_{n 0}^{\dagger}
             \gamma_5 \Gamma^{\dagger} \right] \right\>
    \left\<  \Tr [ \ddot{D}P ] - \Tr [ \dot{D}P\dot{D}P ]
      \right\>  \right\} .
\nonumber \\
 & &
\label{vec2nd2}
\end{eqnarray}

\section{Responses for staggered fermions}

For staggered fermions, the determinant factor $\Delta$ is given by
Eq.~(\ref{del}), and this leads to
\begin{eqnarray}
\frac{ \dot{\Delta} }{ \Delta } = \frac{1}{2} \Tr [ \dot{D}P ] ,
\end{eqnarray}
\begin{eqnarray}
\frac{ \ddot{\Delta} }{ \Delta } = \frac{1}{2} \Tr [ \ddot{D}P ]
 - \frac{1}{2} \Tr [ \dot{D}P\dot{D}P ]
 + \frac{1}{4} \Tr [ \dot{D}P ]^2 .
\label{deri12b}
\end{eqnarray}
The final expressions for the second responses are
\begin{widetext}
\begin{eqnarray}
\frac{d^2}{d\hat{\mu}^2} \Re \left\< H(n)H(0)^{\dagger} \right\>
 &=&
   4 \Re \left\< \Tr \left[ (P\dot{D}P\dot{D}P)_{n 0} \Gamma \gamma_5
          P_{n 0}^{\dagger} \gamma_5 \Gamma^{\dagger} \right] \right\>
  -2 \Re \left\< \Tr \left[ (P\ddot{D}P)_{n 0} \Gamma \gamma_5
          P_{n 0}^{\dagger} \gamma_5 \Gamma^{\dagger} \right] \right\>
\nonumber \\
 & &  \hspace{-3cm}
  -2 \Re \left\< \Tr \left[ (P\dot{D}P)_{n 0}\Gamma \gamma_5
        (P\dot{D}P)_{n 0}^{\dagger} \gamma_5 \Gamma^{\dagger}
                                               \right] \right\>
  +2 \left\< \Im \Tr \left[ (P\dot{D}P)_{n 0}\Gamma \gamma_5
         P_{n 0}^{\dagger} \gamma_5 \Gamma^{\dagger} \right]
                                    \Im \Tr [\dot{D}P] \right\>
\nonumber \\
 & & \hspace{-3cm}
   + \frac{1}{2} \Re \left\{ \;\;
    \left\<  \Tr \left[ P_{n 0}\Gamma \gamma_5 P_{n 0}^{\dagger}
         \gamma_5 \Gamma^{\dagger} \right]
    \left( \Tr [\ddot{D}P] - \Tr [ \dot{D}P\dot{D}P ]
            +\frac{1}{2}\Tr [ \dot{D}P ]^2 \right) \right\> \right.
\nonumber \\
& & \left. \hspace{-1.9cm}
  - \left\<  \Tr \left[ P_{n 0} \Gamma \gamma_5 P_{n 0}^{\dagger}
          \gamma_5 \Gamma^{\dagger} \right] \right\>
    \left\< \Tr [ \ddot{D}P ] - \Tr [ \dot{D}P\dot{D}P ]
         + \frac{1}{2} \Tr [ \dot{D}P ]^2  \right\>  \;    \right\}
\label{sca2nd2}
\end{eqnarray}
for the isoscalar chemical potential,  and
\begin{eqnarray}
\frac{d^2}{d\hat{\mu}^2} \Re \left\< H(n)H(0)^{\dagger} \right\>
 &=&
   4 \Re \left\< \Tr \left[ (P\dot{D}P\dot{D}P)_{n 0} \Gamma \gamma_5
          P_{n 0}^{\dagger} \gamma_5 \Gamma^{\dagger} \right] \right\>
  -2 \Re \left\< \Tr \left[ (P\ddot{D}P)_{n 0} \Gamma \gamma_5
          P_{n 0}^{\dagger} \gamma_5 \Gamma^{\dagger} \right] \right\>
\nonumber \\
 & &
  +2 \Re \left\< \Tr \left[ (P\dot{D}P)_{n 0}\Gamma \gamma_5
        (P\dot{D}P)_{n 0}^{\dagger} \gamma_5 \Gamma^{\dagger}
                                               \right] \right\>
\nonumber \\
 & &
   + \frac{1}{2} \Re \left\{ \;\;
    \left\<  \Tr \left[ P_{n 0}\Gamma \gamma_5 P_{n 0}^{\dagger}
         \gamma_5 \Gamma^{\dagger} \right]
    \left( \Tr [\ddot{D}P] - \Tr [ \dot{D}P\dot{D}P ]
                                           \right) \right\> \right.
\nonumber \\
& & \left. \hspace{1.1cm}
  - \left\<  \Tr \left[ P_{n 0} \Gamma \gamma_5 P_{n 0}^{\dagger}
          \gamma_5 \Gamma^{\dagger} \right] \right\>
    \left\< \Tr [ \ddot{D}P ] - \Tr [ \dot{D}P\dot{D}P ]
            \right\>  \;    \right\}
\label{vec2nd2e}
\end{eqnarray}
for the isovector chemical potential.
\end{widetext}


\end{document}